# Journal Name

## ARTICLE TYPE

# Tracking Thermal-Induced Amorphization of a Zeolitic Imidazolate Framework via Synchrotron *In Situ* Far-Infrared Spectroscopy

Matthew R. Ryder,[a,b] Thomas D. Bennett,[c] Chris S. Kelley,[b] Mark D. Frogley,[b] Gianfelice Cinque[b] and Jin-Chong Tan*[a]

**We present the first use of *in situ* far-infrared spectroscopy to analyze the thermal amorphization of a zeolitic imidazolate framework material. We explain the nature of vibrational motion changes during the amorphization process and reveal new insights into the effect that temperature has on the Zn-N tetrahedra.**

Vibrational spectroscopy has been shown to be an excellent method of analyzing framework materials and gaining a better understanding of their molecular structure. The mid-infrared (MIR) region of the vibrational spectra is related to the local characteristic vibrations (e.g. localized bond stretching and bending) and is therefore of limited interest for providing information regarding lattice dynamics of a specific framework. However, there has recently been a strong focus on the framework specific vibrational motions located in the far-infrared (FIR) region (< 700 cm$^{-1}$), which have been shown to reveal the nature of quasi-localized and collective modes.

One group of framework materials exhibiting a particularly rich variety of collective vibrational motions are metal-organic frameworks (MOFs),[1, 2] which are a topical class of hybrid (inorganic-organic) crystalline materials. Their nanoscale pore structures and long-range order have attracted significant scientific and industrial interest, which originates from their wide range of potential applications including, carbon sequestration, photonics and microelectronics, and drug encapsulation and delivery.[3, 4] The diverse structural behavior of MOFs[5, 6] results in the FIR region of the vibrational spectra providing a gateway to understanding the physical behavior and underlying flexibility of these highly promising next-generation functional materials.

We have shown that inelastic neutron scattering (INS) and synchrotron-based FIR spectroscopy, in conjunction with *ab initio* density functional theory (DFT) can be used to explain these framework specific motions.[7] Specifically, we related quasi-local vibrations to the deformation of the organic linkers and Zn-based coordination polyhedra (ZnN$_4$) in zeolitic imidazolate frameworks (ZIFs),[8] and linked them to various physical phenomena, including "gate-opening", "breathing" and shear destabilization modes.[7] We also reported molecular rotational dynamics, trampoline-like vibrational motions, and demonstrated how collective vibrations can be connected to the elasticity and mechanical stability of the framework structure, revealing the origin of auxeticity (negative Poisson's ratio), in the copper paddle-wheel MOF, HKUST-1.[9] Unlike *in situ* MIR spectroscopy,[10, 11] which can be performed via benchtop equipment, detailed characterization of the FIR region *in situ* requires a more sophisticated experimental setup, explaining the relative absence of reports to date.

The structural dynamics of MOFs, at the molecular level, is of increasing importance. For example, phase transitions in ZIF-7 upon exchange of guest molecules and ZIF-8 upon gas adsorption have been ascribed to "breathing" and "gate-opening" mechanisms respectively.[7, 12-15] Another topical

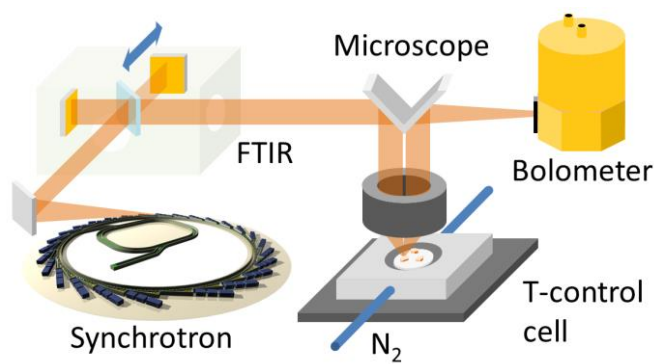

Fig. 1. Schematic of the novel experimental setup used for the *in situ* far-IR spectroscopy study conducted at the B22 Beamline (MIRIAM) at Diamond Light Source.

[a.] Multifunctional Materials and Composites (MMC) Laboratory, Department of Engineering Science, University of Oxford, Parks Road, Oxford OX1 3PJ, United Kingdom. E-mail: jin-chong.tan@eng.ox.ac.uk; Tel: +44 1865 273925
[b.] Diamond Light Source, Harwell Campus, Didcot, Oxford OX11 0DE, United Kingdom.
[c.] Department of Materials Science and Metallurgy, University of Cambridge, Cambridge CB3 0FS, United Kingdom.

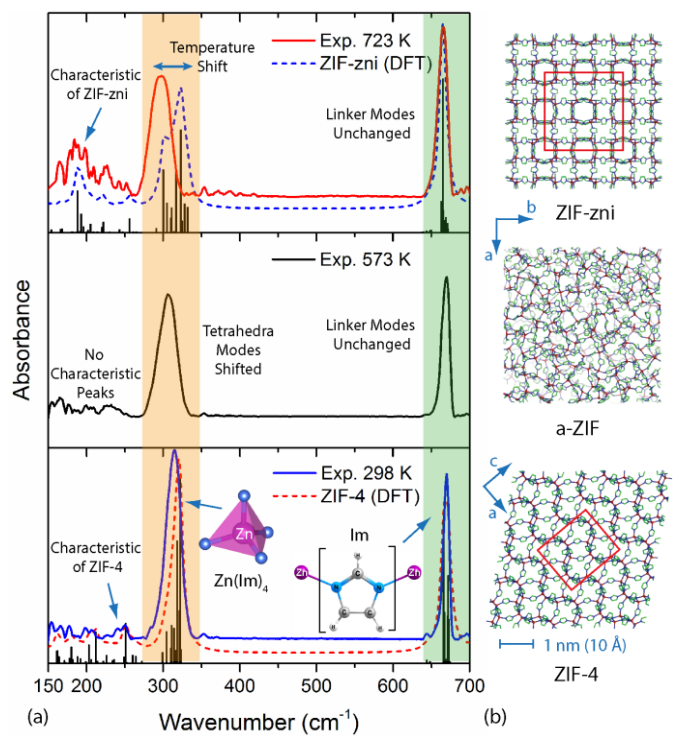

Fig. 2. (a) IR spectra in the region of 150-700 cm$^{-1}$, showing the spectral changes observed at 298, 573 and 723 K. The red and blue dotted lines are the theoretical ZIF-4 and ZIF-zni spectra obtained from DFT (comparable to 0 K). The modes relating to the Zn-N deformations of the Zn(Im)$_4$ tetrahedra and the deformations of the aromatic imidazole (Im) linkers are highlighted. (b) Framework structures adapted from Ref. 19 of ZIF-4, a-ZIF and ZIF-zni are shown for clarity, with the unit cells of the two crystalline phases highlighted in red.

MOF material, MIL-53, has also been reported to undergo transitions upon cooling.[16-18]

In this work, we have advanced the use of FIR vibrational spectroscopy to follow the *in situ* thermal transformations of ZIF-4, which has the chemical composition Zn(Im)$_2$ [Im = imidazolate, $C_3H_3N_2^-$]. The framework was selected, as it has been shown to demonstrate significant flexibility regarding its phase transition ability.[19] It is known to amorphize to a structure known as a-ZIF at approximately 300 °C, which is separate from the solvent induced collapse of ZIFs. Upon further heating (above 400 °C) a-ZIF recrystallizes into the higher symmetry dense structure, ZIF-zni.[19] It has also been reported to undergo a discontinuous porous to dense phase transition upon cooling to under 140 K, due to cooperative rotations of the organic linkers.[20] The temperature ranges of these phases are indicated in Figure 3 by ZIF-4 (LT/HT). For comparison, we have also investigated the thermal effects on the vibrational spectra of ZIF-8 (whose linker is mIm = 2-methylimidazolate), which does not amorphize thermally.[21] It is worth noting that these materials can also amorphize through other sources of external stimuli, such as pressure[22] and mechanical impact.[23] However, it has been suggested that the mechanisms for the various routes to amorphization are different, with a shear instability being the likely cause of the mechanically induced amorphization.[24, 25] The thermal-induced amorphization route of ZIFs is currently a phenomenon only witnessed for ZIF structures possessing the unsubstituted imidazole (Im) linker. We have, therefore, used the results of this work to postulate a reason for this observation and propose in this communication an empirical measure to quantify the structural flexibility determining the likelihood of temperature enabled amorphization.

The experimental spectra were obtained using a novel setup recently commissioned utilizing a Bruker Vertex 80V Fourier Transform IR (FTIR) Interferometer, to provide *in situ* FIR data using a synchrotron radiation source (Figure 1). The novelty of the setup was the introduction of an accurate temperature control, via a nitrogen purged Linkam cell and a Hyperion 3000 microscope.[26] A more detailed description of the experimental setup is available in the Supplementary Information.

We know from our previous work on the low-energy vibrations of ZIF materials[7] that there are three sources of vibrational motion located in the spectral region below 700 cm$^{-1}$:

1. Aromatic rings deformations
2. ZnN$_4$ tetrahedra deformations
3. Framework-specific collective motions

We have confirmed our previous work, demonstrating that the peaks relating to the ring deformations of the Im organic linkers in ZIF-4, a-ZIF and ZIF-zni are all present at approximately 640-700 cm$^{-1}$, along with the universally shared Zn-N bond stretching motion, specifically originating from the compliance of Zn(Im)$_4$ tetrahedra, which is again positioned at 265-325 cm$^{-1}$ as expected (Figure 2).[7]

Unsurprisingly we do not see any significant effect on the peaks relating to the organic linkers, as the structurally rigid Im linkers do not strain significantly upon heating, structural amorphization, or recrystallization into ZIF-zni. Changes to the peaks at approximately 323 cm$^{-1}$ (assuming zero thermal effects, calculated via DFT) are however pronounced and include a temperature induced spectral shift and are related to the Zn-N stretching modes of the ZnN$_4$ tetrahedra deformation in Zn(Im)$_4$. These vibrational modes are expected to be sufficiently anharmonic to show a red shift as a result of the thermal stimulus (upon heating) allowing for increased flexibility in the deformation of the Zn(Im)$_4$ tetrahedra. Due to these modes (phonons) being bosons and present in each of the phases, we expect these peaks to obey a Bose-Einstein distribution.[27, 28] The frequency shift (Figure 3) can, therefore, be explained by the following equation:

$$\nu(T) = \nu_0 - \frac{X_R T_C}{e^{\frac{T_C}{T}} - 1}$$

$$T_C = \frac{h\nu_0}{k_B}$$

where $\nu_0$ is the frequency of the mode at zero temperature (calculated via DFT); $T$ is the temperature that the experimental spectra were obtained at; $T_C$ is the vibrational temperature, related to the energy of the particular mode, obtained from statistical thermodynamics, via the use of the Planck constant ($h$) and the Boltzmann constant ($k_B$); and $X_R$ is the thermally-induced spectral shift coefficient, which we postulate is unique to each specific M-X polyhedral deformation, present in all MOF structures. The value of $X_R$ is dependent on both the type of metal and importantly the linker involved and is 0.045 cm$^{-1}$ K$^{-1}$ for Zn(Im)$_4$ tetrahedral deformations in Im containing ZIFs, with an adjusted R$^2$ value of 0.970 for the curve fitting (explained in detail in the Supplementary Information). We, therefore, postulate that MOF structures that demonstrate a higher value of $X_R$ will be more susceptible to thermal-induced amorphization. To strengthen this claim, we investigated the value of $X_R$ for ZIF-8, which is well accepted not to amorphize thermally.[21] Importantly, the linker is different, 2-methylimidazole (mIm), so we could provide another value of $X_R$, this time applied to Zn(mIm)$_4$ tetrahedral deformations. The resultant value for ZIF-8 is 0.032 cm$^{-1}$ K$^{-1}$, with an adjusted R$^2$ value of 0.995 for the curve fitting, hence prompting the claim that a lower value of $X_R$ implies an increased thermal stability to amorphization. It is important to note that in the literature, the flexibility of the Zn-N tetrahedra has generally been discarded as a major factor in further understanding the mechanisms responsible for the amorphization of ZIFs[29] and this current work reveals that there is still much that can be learned from analyzing these important tetrahedral motions.

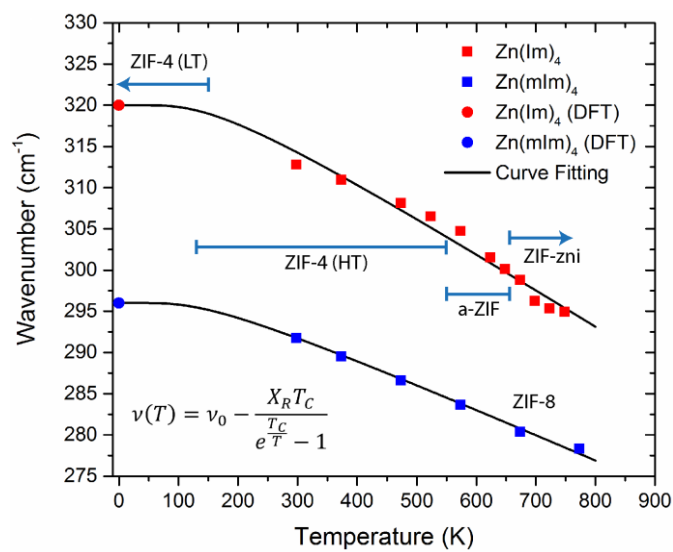

Fig. 3. Temperature dependence of the Zn(Im)$_4$ and Zn(mIm)$_4$ tetrahedra deformation modes, hence explaining the disparity between the theoretical ZIF-zni spectra and the experimental spectra obtained at 723 K. The thermal range of each structure has been highlighted for clarity.

Table 1. Nature of the IR-active modes in the 150-700 cm$^{-1}$ region.

| Structure | Spectral Region (cm$^{-1}$) | Description of Vibrational Motion |
|---|---|---|
| ZIF-4 | 160-275 | 4-Membered and 6-Membered Ring Deformation; Ligand Rotation (Rocking); N-Zn-N Bending and Zn-N Stretching (Tetrahedral Deformation) |
| ZIF-zni | 155-280 | |
| ZIF-4 | 295-325 | Zn-N Stretching (Belonging to ZnN$_4$ Tetrahedra) |
| ZIF-zni | 290-330 | |
| ZIF-4 and ZIF-zni | 640-700 | Aromatic Ring Deformation and Torsion Modes (In-Plane and Out-of-Plane) |

Moving on to the distinction between the two crystalline structures, the final modes observed are the collective vibrational motions that are specific to the framework structure and symmetry. These modes are located in the region of 155-280 cm$^{-1}$ (Figure 2) and are all due to specific motions involving the 4-membered and 6-membered ring deformations, simultaneously coupled with ligand rocking.[7] Due to the differences in crystal packing and symmetry between ZIF-4 and ZIF-zni the intensities of the peaks in this region are distinct to each framework. For example, in ZIF-4 we observe a broader distribution of modes due to the lower symmetry of the framework, with the highest in intensity being located at approximately 250 cm$^{-1}$. In ZIF-zni, there are more degenerate modes, and therefore there is a significant intensity at approximately 195 cm$^{-1}$. This region shows less significant features at 300 °C, as the amorphous nature of the framework will negate any structural symmetry and result in such motions being so widely (randomly) distributed that no identifiable characteristic intensities are observed.

For completeness, we also analyzed the mechanical properties of ZIF-zni (Table 2), for comparison to our previous elasticity work on the other Im containing ZIFs (ZIF-1, -2, -3 and -4).[25] We performed the theoretical calculations at the same B3LYP level of theory[30-32] as our previous work (explained in detail in the Supplementary Information), so as to be able to comment on the quantitative mechanical property differences. The revised calculations reported in this communication and our recent work on the other four Im containing ZIF structures[25] are an improved version to our initial calculations performed using the PBE level of theory.[33] Because of the higher density and increased symmetry of the ZIF-zni structure compared with ZIF-4, we observe an increase in the framework stiffness of the material (Young's moduli) by a factor of ~3. In addition, there is a two-fold increase in the resistance to shear strain through doubling of maximum shear modulus ($G_{max}$), which can contribute to the mechanical destabilization of ZIF materials.[7,]

Table 2. Mechanical properties of ZIF-4 and ZIF-zni, derived from the elastic constants calculated by DFT at the B3LYP level of theory.

| Elastic Property | | ZIF-4[24] | ZIF-zni |
|---|---|---|---|
| Young's Modulus, $E$ (GPa) | $E_{max}$ | 3.27 | 10.21 |
| | $E_{min}$ | 2.17 | 6.82 |
| | $A_E = E_{max}/E_{min}$ | 1.51 | 1.50 |
| Shear Modulus, $G$ (GPa) | $G_{max}$ | 1.53 | 3.26 |
| | $G_{min}$ | 0.77 | 2.30 |
| | $A_G = G_{max}/G_{min}$ | 1.99 | 1.42 |
| Linear Compressibility, $\beta$ (TPa$^{-1}$) | $\beta_{max}$ | 241.2 | 29.5 |
| | $\beta_{min}$ | 202.1 | 5.2 |
| Poisson's Ratio, $\nu$ | $\nu_{max}$ | 0.41 | 0.49 |
| | $\nu_{min}$ | 0.06 | 0.32 |

25 The Poisson's ratio was also affected, in that it no longer possesses the 'cork-like' response (zero Poisson's ratio), in any direction, as is the case with ZIF-4 whose $\nu_{min} \approx$ zero. The values for the Young's moduli agree with experimental nanoindentation values,[19] with the experimental $E_{max}$ of ZIF-zni in <001> direction ~9 GPa, and the experimental $E_{<100>}$ ~8 GPa (DFT value from this work, $E_{<100>}$ = 8.6 GPa).

In conclusion, the thermally induced amorphization of ZIF-4 into a-ZIF and its subsequent recrystallization upon additional heating to transform into ZIF-zni has been followed *in situ* using synchrotron Far-IR spectroscopy for the first time. We have demonstrated how vibrational spectroscopy can be used to gain a better understanding of how thermal stimulus can affect the stability of framework materials. Our work confirms the nature of each different vibrational motion and makes the connection between the flexibility (mechanical compliance) of the Zn-N tetrahedral moieties and the likelihood of thermal-induced amorphization. Also, we propose a new relationship that offers a quantitative measure of the thermal stability to amorphization via the magnitude of the coefficient $X_R$ and encourage future work in this area to test its wider applicability. In relation, we would expect other ZIF structures that possess the unsubstituted Im organic linker (e.g. ZIF-1, -2, -3, and -10) to have a similar value of $X_R$ to that reported for ZIF-4, a-ZIF, and ZIF-zni in this work ($X_R$ = 0.045 cm$^{-1}$ K$^{-1}$). It is also of great interest to further this work into MOFs with different metal polyhedra coordination, such as the high stability UiO series[34] and the isoreticular MIL-140 series.[35, 36]

M. R. R. would like to thank the UK Engineering and Physical Sciences Research Council (EPSRC) for a DTA postgraduate scholarship and an additional scholarship from the Science and Technology Facilities Council (STFC) CMSD Award 13-05. The experimental work was performed at the large scale facility Diamond through beamtime (SM10215) at the B22 MIRIAM beamline. We thank the Advanced Research Computing (ARC) facility (http://dx.doi.org/10.5281/zenodo.22558) at Oxford University and the SCARF cluster at the Rutherford Appleton Laboratory for additional computing resources. T. D. B. would like to thank the Royal Society for funding.

## Notes and references

1. H. Furukawa, K. E. Cordova, M. O'Keeffe and O. M. Yaghi, *Science*, 2013, **341**, 1230444.
2. H. C. Zhou, J. R. Long and O. M. Yaghi, *Chem. Rev.*, 2012, **112**, 673-674.
3. M. R. Ryder and J. C. Tan, *Mater. Sci. Technol.*, 2014, **30**, 1598-1612.
4. C. Y. Sun, C. Qin, X. L. Wang and Z. M. Su, *Expert Opin. Drug Deliv.*, 2013, **10**, 89-101.
5. L. Sarkisov, R. L. Martin, M. Haranczyk and B. Smit, *J. Am. Chem. Soc.*, 2014, **136**, 2228-2231.
6. A. Schneemann, V. Bon, I. Schwedler, I. Senkovska, S. Kaskel and R. A. Fischer, *Chem. Soc. Rev.*, 2014, **43**, 6062-6096.
7. M. R. Ryder, B. Civalleri, T. D. Bennett, S. Henke, S. Rudic, G. Cinque, F. Fernandez-Alonso and J. C. Tan, *Phys. Rev. Lett.*, 2014, **113**, 215502.
8. K. S. Park, Z. Ni, A. P. Cote, J. Y. Choi, R. Huang, F. J. Uribe-Romo, H. K. Chae, M. O'Keeffe and O. M. Yaghi, *Proc. Natl. Acad. Sci. U. S. A.*, 2006, **103**, 10186-10191.
9. M. R. Ryder, B. Civalleri, G. Cinque and J. C. Tan, *CrystEngComm*, 2016, **18**, 4303-4312.
10. Z. H. Dong and Y. Song, *J. Phys. Chem. C*, 2010, **114**, 1782-1788.
11. A. Greenaway, B. Gonzalez-Santiago, P. M. Donaldson, M. D. Frogley, G. Cinque, J. Sotelo, S. Moggach, E. Shiko, S. Brandani, R. F. Howe and P. A. Wright, *Angew. Chem., Int. Ed. Engl.*, 2014, **53**, 13483-13487.
12. C. Gucuyener, J. van den Bergh, J. Gascon and F. Kapteijn, *J. Am. Chem. Soc.*, 2010, **132**, 17704-17706.
13. S. Aguado, G. Bergeret, M. P. Titus, V. Moizan, C. Nieto-Draghi, N. Bats and D. Farrusseng, *New J. Chem.*, 2011, **35**, 546-550.
14. P. Zhao, G. I. Lamperti, G. O. Lloyd, M. T. Wharmby, S. Facq, A. K. Cheetham and S. A. Redfern, *Chem. Mater.*, 2014, **26**, 1767-1769.
15. S. A. Moggach, T. D. Bennett and A. K. Cheetham, *Angew. Chem., Int. Ed. Engl.*, 2009, **48**, 7087-7089.
16. C. Serre, S. Bourrelly, A. Vimont, N. A. Ramsahye, G. Maurin, P. L. Llewellyn, M. Daturi, Y. Filinchuk, O. Leynaud, P. Barnes and G. Ferey, *Adv. Mater.*, 2007, **19**, 2246-2251.
17. Y. Liu, J. H. Her, A. Dailly, A. J. Ramirez-Cuesta, D. A. Neumann and C. M. Brown, *J. Am. Chem. Soc.*, 2008, **130**, 11813-11818.
18. A. Ghoufi, A. Subercaze, Q. Ma, P. G. Yot, Y. Ke, I. Puente-Orench, T. Devic, V. Guillerm, C. Zhong, C. Serre, G. Ferey and G. Maurin, *J. Phys. Chem. C*, 2012, **116**, 13289-13295.
19. T. D. Bennett, A. L. Goodwin, M. T. Dove, D. A. Keen, M. G. Tucker, E. R. Barney, A. K. Soper, E. G. Bithell, J. C. Tan and A. K. Cheetham, *Phys. Rev. Lett.*, 2010, **104**, 115503.
20. M. T. Wharmby, S. Henke, T. D. Bennett, S. R. Bajpe, I. Schwedler, S. P. Thompson, F. Gozzo, P. Simoncic, C. Mellot-Draznieks, H. Tao, Y. Yue and A. K. Cheetham, *Angew. Chem., Int. Ed. Engl.*, 2015, **54**, 6447-6451.
21. T. D. Bennett, J. C. Tan, Y. Yue, E. Baxter, C. Ducati, N. J. Terrill, H. H. Yeung, Z. Zhou, W. Chen, S. Henke, A. K. Cheetham and G. N. Greaves, *Nat. Comm.*, 2015, **6**, 8079.
22. T. D. Bennett, P. Simoncic, S. A. Moggach, F. Gozzo, P. Macchi, D. A. Keen, J. C. Tan and A. K. Cheetham, *Chem. Commun.*, 2011, **47**, 7983-7985.
23. B. Van de Voorde, I. Stassen, B. Bueken, F. Vermoortele, D. De Vos, R. Ameloot, J.-C. Tan and T. D. Bennett, *J. Mater. Chem. A*, 2015, **3**, 1737-1742.
24. A. U. Ortiz, A. Boutin, A. H. Fuchs and F. X. Coudert, *J. Phys. Chem. Lett.*, 2013, **4**, 1861-1865.
25. M. R. Ryder and J. C. Tan, *Dalton Trans.*, 2016, **45**, 4154-4161.
26. G. Cinque, M. D. Frogley, K. Wehbe, J. Filik and J. Pikjanka, *Synchrotron Radiation News*, 2011, **24**, 24-33.
27. L. Vina, S. Logothetidis and M. Cardona, *Phys. Rev. B*, 1984, **30**, 1979-1991.
28. Y. C. Shen, P. C. Upadhya, E. H. Linfield and A. G. Davies, *Appl. Phys. Lett.*, 2003, **82**, 2350-2352.
29. E. O. Beake, M. T. Dove, A. E. Phillips, D. A. Keen, M. G. Tucker, A. L. Goodwin, T. D. Bennett and A. K. Cheetham, *J. Phys.: Condens. Matter*, 2013, **25**, 395403.
30. A. D. Becke, *J. Chem. Phys.*, 1993, **98**, 5648-5652.
31. C. T. Lee, W. T. Yang and R. G. Parr, *Phys. Rev. B*, 1988, **37**, 785-789.
32. P. J. Stephens, F. J. Devlin, C. F. Chabalowski and M. J. Frisch, *J. Phys. Chem.*, 1994, **98**, 11623-11627.
33. J. C. Tan, B. Civalleri, A. Erba and E. Albanese, *CrystEngComm*, 2015, **17**, 375-382.

34. J. H. Cavka, S. Jakobsen, U. Olsbye, N. Guillou, C. Lamberti, S. Bordiga and K. P. Lillerud, *J. Am. Chem. Soc.*, 2008, **130**, 13850-13851.
35. M. R. Ryder, B. Civalleri and J. C. Tan, *Phys. Chem. Chem. Phys.*, 2016, **18**, 9079-9087.
36. V. Guillerm, F. Ragon, M. Dan-Hardi, T. Devic, M. Vishnuvarthan, B. Campo, A. Vimont, G. Clet, Q. Yang, G. Maurin, G. Ferey, A. Vittadini, S. Gross and C. Serre, *Angew. Chem., Int. Ed. Engl.*, 2012, **51**, 9267-9271.